\documentclass{article}
\usepackage[paper=a4paper, dvips, top=1.5cm, left=1.5cm, right=1.5cm, foot=1.5cm, bottom=1.5cm]{geometry}
\usepackage[table]{xcolor}
\usepackage{cite,url}
\usepackage{setspace}
\usepackage{subcaption}
\usepackage{graphicx}
\usepackage{float}
\usepackage{booktabs}
\usepackage{amsmath} 
\usepackage{amssymb}  
\usepackage[normalem]{ulem}
\usepackage{hyperref}

{}
{}
{}

\begin{document}
	\title{Period-Amplitude Co-variation in Biomolecular Oscillators}
	\author{{Venkat Bokka$^{1}$}, {Abhishek Dey$^{2}$}, {Shaunak Sen$^{3}$}\\
		Department of Electrical Engineering,\\Indian Institute of Technology Delhi\\
		Hauz Khas, New Delhi 110016, INDIA\\
		E-mail: venkatb@ee.iitd.ac.in$^{1}$, abhishek.dey@ee.iitd.ac.in$^{2}$, shaunak.sen@ee.iitd.ac.in$^{3}$
	}
	
	\date{\color{blue}Accepted in IET Systems Biology. DOI: \href{http://dx.doi.org/10.1049/iet-syb.2018.0015}{10.1049/iet-syb.2018.0015}}
	
	\maketitle

\begin{abstract}
The period and amplitude of biomolecular oscillators are functionally important properties in multiple contexts.
For a biomolecular oscillator, the overall constraints in how tuning of amplitude affects period, and vice versa, are generally unclear.
Here we investigate this co-variation of the period and amplitude in mathematical models of biomolecular oscillators using both simulations and analytical approximations.
We computed the amplitude-period co-variation of eleven benchmark biomolecular oscillators as their parameters were individually varied 
around a nominal value, classifying the various co-variation patterns such as a simultaneous increase/ decrease in period and amplitude. Next, we repeated the classification using a power norm-based amplitude metric, to account for the amplitudes of the many biomolecular species that may be part of the oscillations, finding largely similar trends. Finally, we calculate "scaling laws" of period-amplitude co-variation for a subset of these benchmark 
oscillators finding that as the approximated period increases, the upper bound of the amplitude increases, or reaches a constant value. Based on these results, we discuss the effect of different parameters on the type of period-amplitude co-variation as well as the difficulty in achieving an oscillation with large amplitude and small period.
\end{abstract}

\maketitle

\section{Introduction}\label{sec1}

Oscillatory behaviour in biomolecular systems span multiple scales of space and time.
An important example is that of circadian rhythms, which are  present in both eukaryotes and prokaryotes.
The classical studies of the genes \textit{period} and \textit{timeless} in \textit{Drosophila melanogaster} provided a genetic context to these oscillations~\cite{BEAVER20041492}.
More recently, biochemical investigations have proved invaluable in investigating the interactions between the proteins KaiABC that are at the core of the cyanobacterial circadian clock~\cite{Nakajima414,Rust809}.
In addition to these examples of naturally occurring oscillatory behaviour, significant effort has been devoted to designing oscillators for synthetic biology applications inside cells and in biochemical contexts.
Examples of synthetic oscillators include the repressilator~\cite{elowitz} and Smolen oscillator~\cite{Smolen6644} in cells, the repressilator and its variations in cell extracts~\cite{HenrikeeLife09771}, as well as oscillators \textit{in vitro}~\cite{jkim, kim2011synthetic, chen2013programmable} (please see~\cite{Purcell1503} for more examples).
In these investigations of analysis and design of oscillators, mathematical models have played an important role in understanding the underlying principles.

An important guiding principle for oscillator investigations have been the presence of a negative feedback with an additional delay-causing mechanism, either explicitly through a time delay such as due to one or more intermediate steps or due to a positive feedback~\cite{Pomerening2005565,Boris,Forger11012005,ananth}. Systems-level studies on these oscillators have characterized the extent of robustness of different mechanistic realizations of oscillations~\cite{strogatz,hart}.
These have also addressed various aspects of oscillator function.
For example, a recent study showed that the amplitude-period co-variation in an oscillator can depend on underlying mechanisms~\cite{tsai}.
In particular, it was argued that a combination of positive and negative feedback enables a broad range of periods for a fixed amplitude, in contrast to mechanisms based on negative feedback alone.
These studies represent important work towards connecting the structure and function of natural as well as synthetic biomolecular oscillators.

\begin{figure}[t]
	\centering
	\includegraphics[scale=0.4]{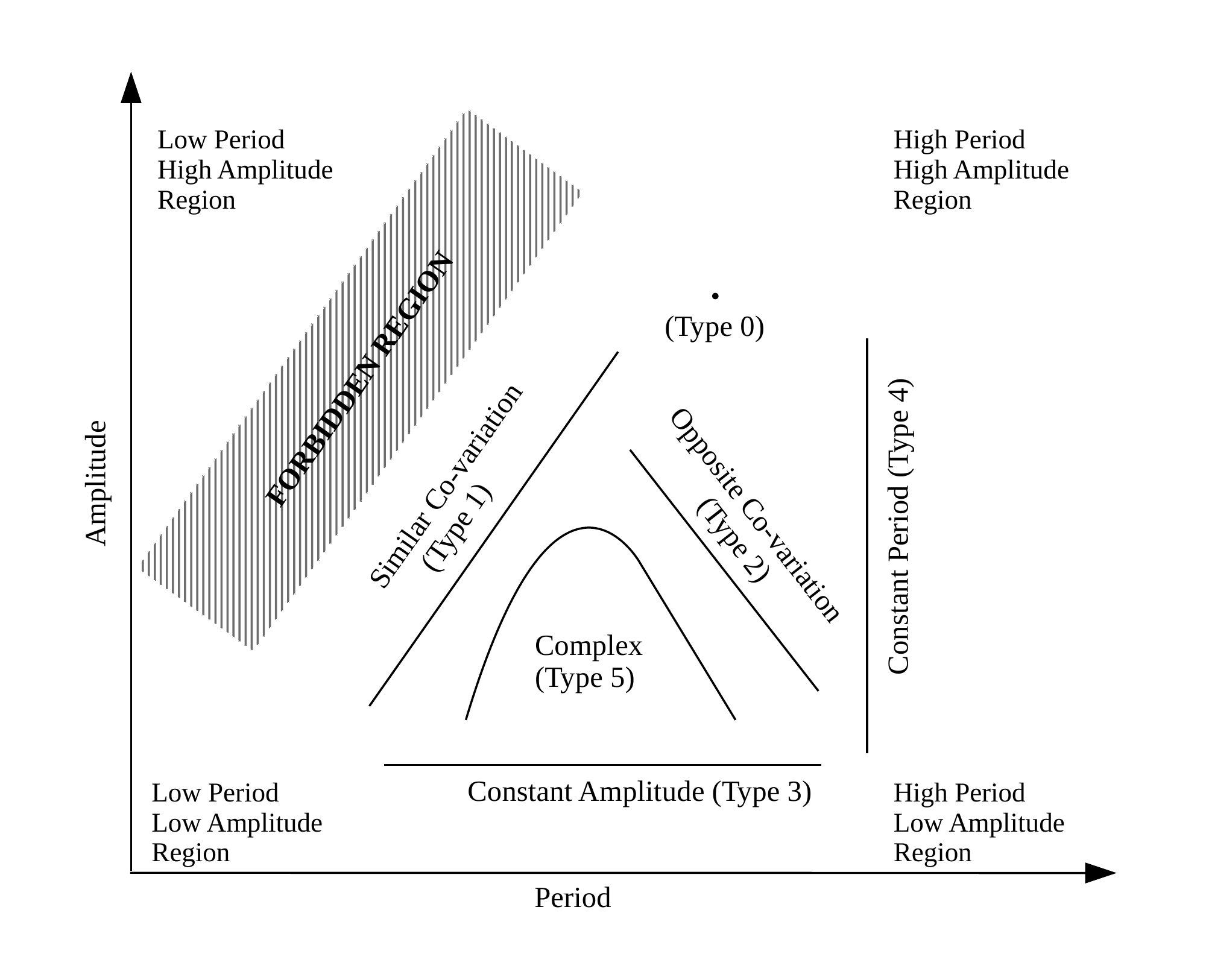}
	\vspace{-.15in}
	\caption{Possible period-amplitude co-variations in biomolecular oscillators categorized as Type 0: both amplitude and period constant/ fixed,  Type 1: similar co-variation, Type 2: opposite co-variation, Type 3: constant amplitude, Type 4: constant period and Type 5: complex.}\label{fig1}
	\vspace{-.15in}
\end{figure}

There are at least three interesting aspects of the amplitude-period co-variation in oscillators.
First, as mentioned above, the amplitude-period co-variation has different qualitative characteristics depending on whether the oscillator has both positive and negative feedback or only negative feedback~\cite{tsai}.
Second, an experimental study noted the correlation between amplitude and period~\cite{jkim}.
Third, from energetic considerations related to protein production and degradation, it seems less favourable to have both large amplitude and small period.
Given these, whether there are constraints and/or patterns in the amplitude-period co-variation in biomolecular oscillator models is generally unclear (Fig. \ref{fig1}).

Here, we investigate the period-amplitude co-variations in biomolecular oscillators. For this, we use mathematical models of an array of benchmark biomolecular oscillators --- Repressilator, Pentilator, Goodwin oscillator, Van der Pol oscillator, Fitzhugh-Nagumo oscillator, Frzilator, Cyanobacterial circadian oscillator, Metabolator, a mixed feedback oscillator, Meyer-Stryer model of calcium oscillations and Kim-Forger oscillator --- and compute the period and amplitude of trajectories for different values of parameters around a nominal parameter set, varying parameters one at a time. We find that the co-variation of the period and amplitude can be similar (Type 1 in Fig. \ref{fig1}), opposite (Type 2), independent where one of the properties is a constant while other varies (Types 3 and 4), or 

\noindent exhibit a combination of these trends (Type 5). To account for multiple possible biomolecular species in an oscillator, we quantify amplitude using a power norm, finding largely similar trends. Finally, we analytically approximate a subset of the oscillator models to calculate "scaling laws" for the period-amplitude co-variation, finding that as the approximate period increases, the upper bound of the amplitude increases or reaches a constant (Type 1/ Type 3). These results provide insight into the period-amplitude co-variation of biomolecular oscillators as well as a reference for parameters that can be tuned to achieve desired amplitude and frequency.

\section{Results}\label{sec2}
\subsection{Simulation of Oscillations}

We considered previously developed mathematical models of eleven benchmark 
oscillators --- Repressilator~\cite{elowitz}, Pentilator~\cite{tsai}, Goodwin oscillator~\cite{goodwin}, Van der Pol oscillator~\cite{vander}, Fitzhugh-Nagumo oscillator~\cite{fitz,nagumo}, Frzilator~\cite{Igoshin02112004},  Cyanobacterial circadian oscillator~\cite{Nakajima414,Rust809}, Metabolator~\cite{Fung}, Mixed feedback oscillator~\cite{hasty, stricker}, Meyer-Stryer model of calcium oscillations~\cite{Meyer01071988} and Kim-Forger model ~\cite{kimforger}. Although Van der Pol oscillator  is not a biomolecular oscillator, we note that it is a benchmark nonlinear oscillator and corresponding period-amplitude co-variation is worth investigating.
These models are based on ordinary differential equations (Please see Supplementary material). To obtain the oscillatory trajectories, these equations were simulated using the MATLAB solvers \textit{ode15s} and \textit{ode23t}.

\indent For each oscillator, we plotted the trajectories by varying one parameter at a time over a twofold range around the nominal value.
These are shown in a red-blue colour map with the X-axis representing time, Y-axis representing parameter value, and the colour representing the instantaneous concentration level along the time axis. We also show oscillatory trajectories picked for three values in the parameter regime (Please see Supplementary material : Figs. 1-11 A).

\subsection{Co-variation of Period and Maximum Amplitude}
We begin by computing the co-variation of the period and maximum amplitude. To find the period and amplitude, we first numerically confirm the existence of oscillations. We define the maximum amplitude as the absolute value of the difference between the maximum/peak and the minimum/bottom of a periodically oscillating trajectory of concentration levels. The period is calculated from the average time-interval between five crests. 
These results are plotted in Supplementary material (Supplementary Figs. 1-11 B).
A zoomed in version of these trends are shown in Fig. \ref{fig2}.
The overall trends are summarized in Table \ref{tab1} (Columns 2 and 3 in subtables 1a and 1b). The rows where the maximum amplitude and period exhibit like co-variation are shaded for clarity.

\subsection{Co-variation of Period and a Power Norm-based Amplitude Metric}

In the above computations, we computed the maximum amplitude based on 
one biomolecular species, such as the protein concentration in the 
Repressilator.  The maximum amplitude is a commonly used measure of 
amplitude (for example in \cite{tsai}), which may be due to the direct visual correspondence with the oscillating trajectory. Another measure used, in 
the context of a circadian clock, is the square-root of maximum power in 
the periodogram \cite{deng2016}. A third measure used, in \cite{Kurosawa}, is the geometric 
mean of all biomolecular species amplitudes, although it is noted that 
other measures may be better. We note that most oscillator models have 
more than one biomolecular species and different period-amplitude 
co-variations may exist for different biomolecular species. To take 
these into account, we consider an amplitude metric based on a power 
norm ~\cite{doyle}, which is similar to the one used in \cite{deng2016}.

\indent Consider a limit-cycle oscillatory system $\mathbf{\dot x}(t) = f(\mathbf{x},p)$, where states $\mathbf{x}\in \Re^n$ and parameter set $p \in \Re^m$, then we define the power norm in one-cycle duration of $T$ as,
 
\begin{equation} \label{metric}
\begin{aligned}
M_p  & = \sqrt{\dfrac{1}{T}\int\limits_{0}^{T} [\sum_{i=1}^{n} x^2_i(t)] dt},
\end{aligned}
\end{equation}

\noindent where $x_i$ are the states in differential equation of the oscillator model.
Based on this definition, we computed the co-variation of the period and this amplitude metric.
The full simulations are shown in Supplementary (Supplementary Figs. 1-11 C).
A zoomed in version is shown in Fig. \ref{fig3}. The results of these calculations are also summarized in Table \ref{tab1} 
(columns 3 and 4 in subtables 1a and 1b) . As before, the rows where the period and this 
amplitude metric exhibit like co-variation are shaded for clarity.

These results provide a classification of period-amplitude co-variations in different biomolecular oscillators.

\subsection{Co-variation scaling laws}
In the above sections we have computed the co-variation of the period 
and amplitude metrics for different benchmark oscillators as individual 
parameters are varied.
Next, we investigate this aspect further by focusing on particular 
oscillators to get analytical approximations.

\begin{figure}[!t]
	\includegraphics[scale=0.8]{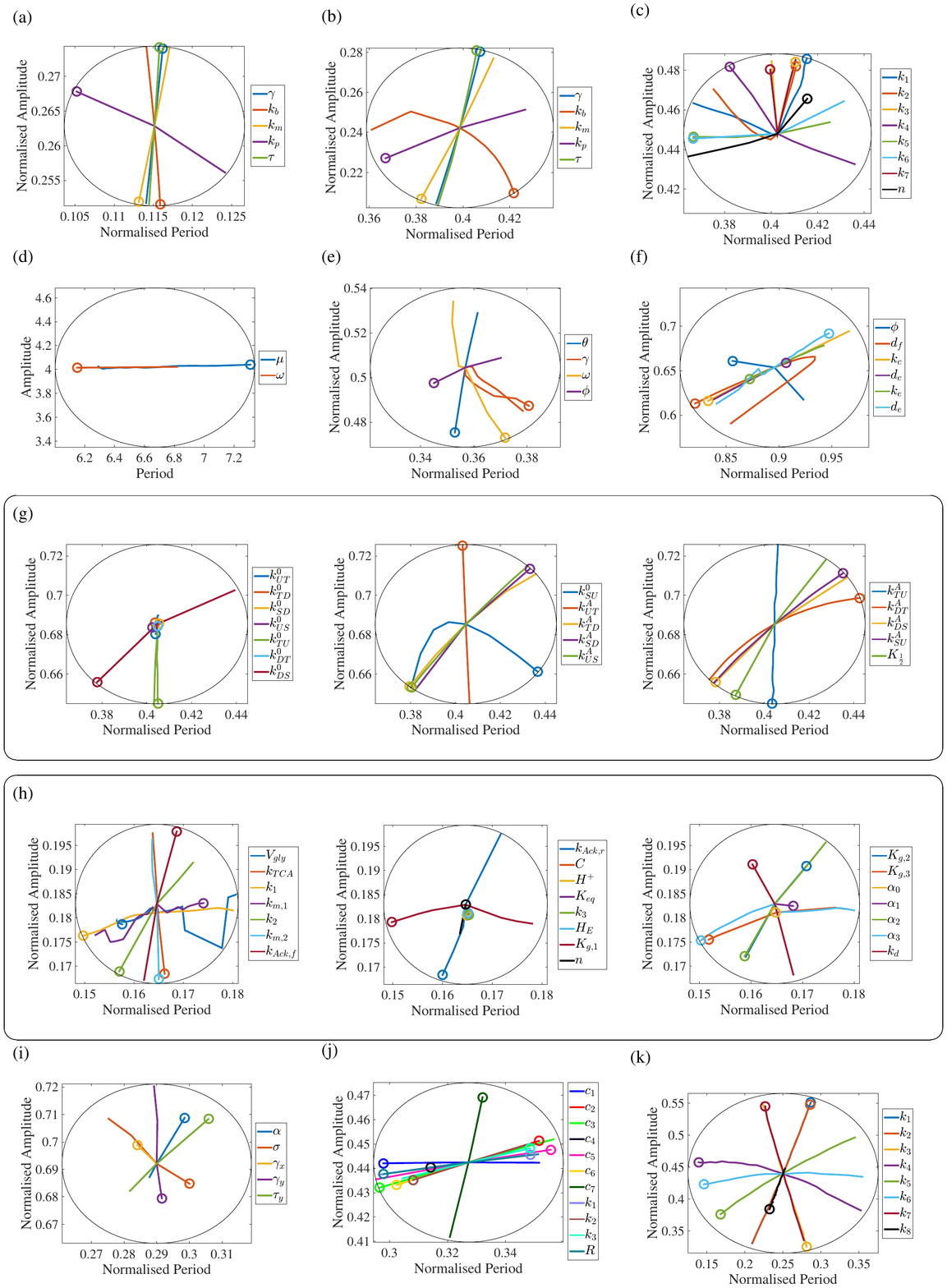}
	\caption{Co-variation of period and maximum amplitude for biomolecular oscillators. In each plot, the point of intersection of the curves is the center of the black circle which corresponds to the nominal parameter set and the radius  is $10\%$ of oscillation period for nominal parameter set. Circle shaped markers represent the largest value of the corresponding parameter. a) Repressilator. b) Pentilator. c) Goodwin oscillator. d) Van der Pol oscillator. e) Fitzhugh-Nagumo oscillator. f) Frzilator. g) Cyanobacteria circadian oscillator. h) Metabolator. i) Mixed feedback oscillator. j) Meyer-Stryer model of calcium oscillations. k) Kim-Forger model.}\label{fig2}
\end{figure}

\begin{figure}[H]
	\centering
	\includegraphics[scale=0.8]{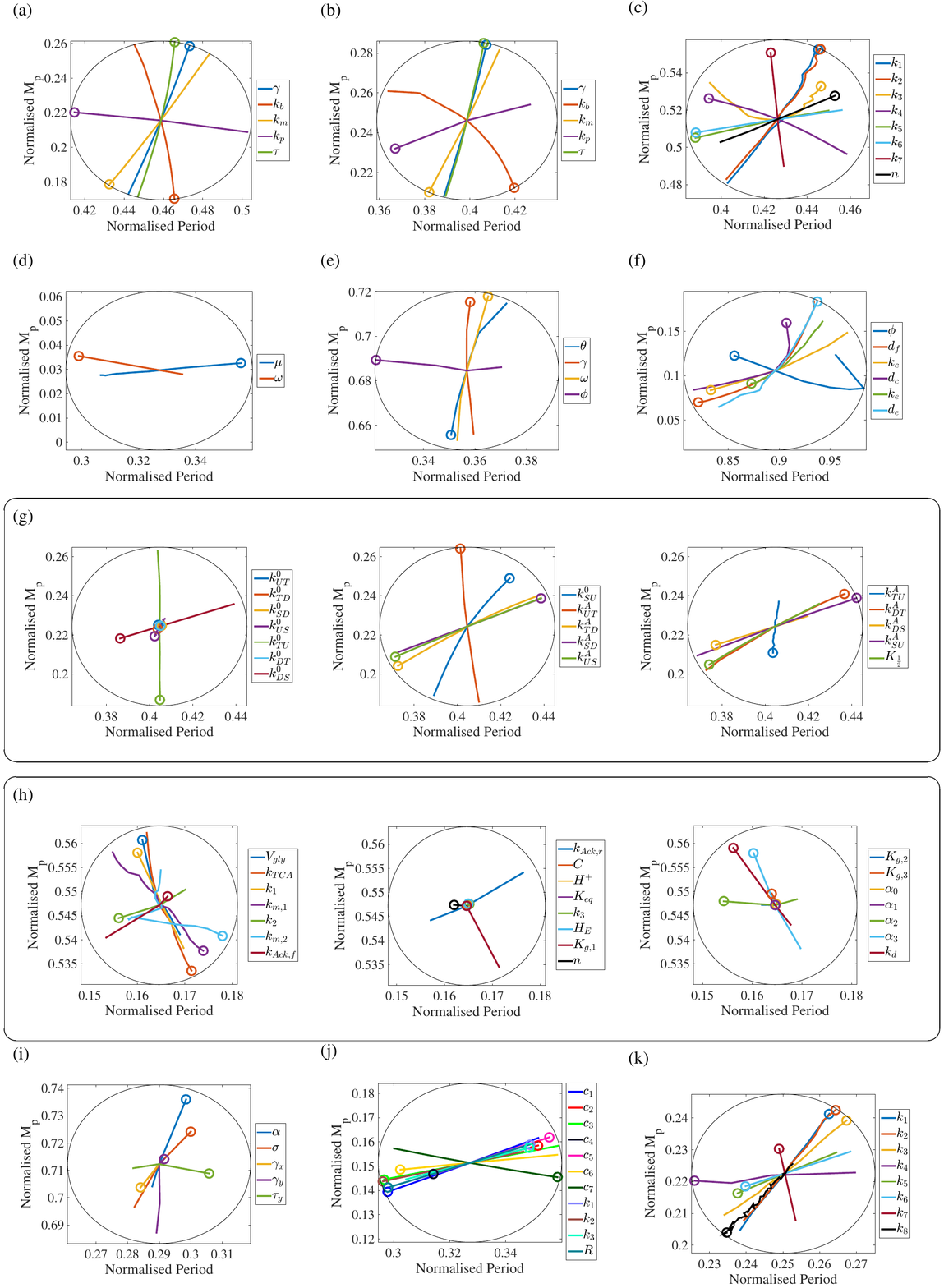}
	\caption{Co-variation of period and amplitude metric ($M_p$) for biomolecular oscillators. In each plot, the point of intersection of the curves is the center of the black circle which corresponds to the nominal parameter set and the radius  is $10\%$ of oscillation period for nominal parameter set. Circle shaped markers represent the largest value of the corresponding parameter. a) Repressilator. b) Pentilator. c) Goodwin oscillator. d) Van der Pol oscillator. e) Fitzhugh-Nagumo oscillator. f) Frzilator. g) Cyanobacteria circadian oscillator. h) Metabolator. i) Mixed feedback oscillator. j) Meyer-Stryer model of calcium oscillations. k) Kim-Forger model.}\label{fig3}
\end{figure}

Consider the Repressilator, one of the most important benchmark for 
oscillator designs.
As parameters are varied individually around their default values, the 
co-variation of period and maximum amplitude take on different 
shapes. For example, for the maximal production rate $\gamma$, both amplitude and 
period increase simultaneously (Fig. \ref{fig2}a). For the protein degradation constant $k_p$, the shape is more complicated. In general, there may be bounds such as on the maximum achievable amplitude for a given period.
To understand this further, we used a framework for analytical approximations developed where a nonlinear oscillatory system is 
replaced by a sequence of linear operations for different parts of the oscillation cycle \cite{Kut2009}.

For the Repressilator, the approximations to period ($T_{approx}$) and amplitude ($A_{approx}$) are given by,
\setlength{\abovedisplayskip}{3pt}%
\setlength{\belowdisplayskip}{3pt}%
\setlength{\abovedisplayshortskip}{3pt}%
\setlength{\belowdisplayshortskip}{3pt}%
\setlength{\jot}{3pt}
\begin{equation}\label{repress_analytic}
	\begin{aligned}
		T_{approx}  & = 3\Big[2 \dfrac{\ln(2)}{k_m}+\dfrac{1}{k_p}\Big(\ln(X)+\ln\Big(\dfrac{X}{X-1}\Big)\Big)\Big], \\
		A_{approx}  & = \dfrac{\tau \gamma}{k_p k_m},
	\end{aligned}
\end{equation} where $X = \dfrac{\tau \gamma \sqrt{k_b}}{k_p k_m}$. 
\\ The $A_{approx}$ is the maximal possible value of the protein 
concentration. As the minimum value is zero. $A_{approx}$ is the upper bound on the maximum amplitude.

\begin{figure}[H]
	\centering
	\includegraphics[scale=.75]{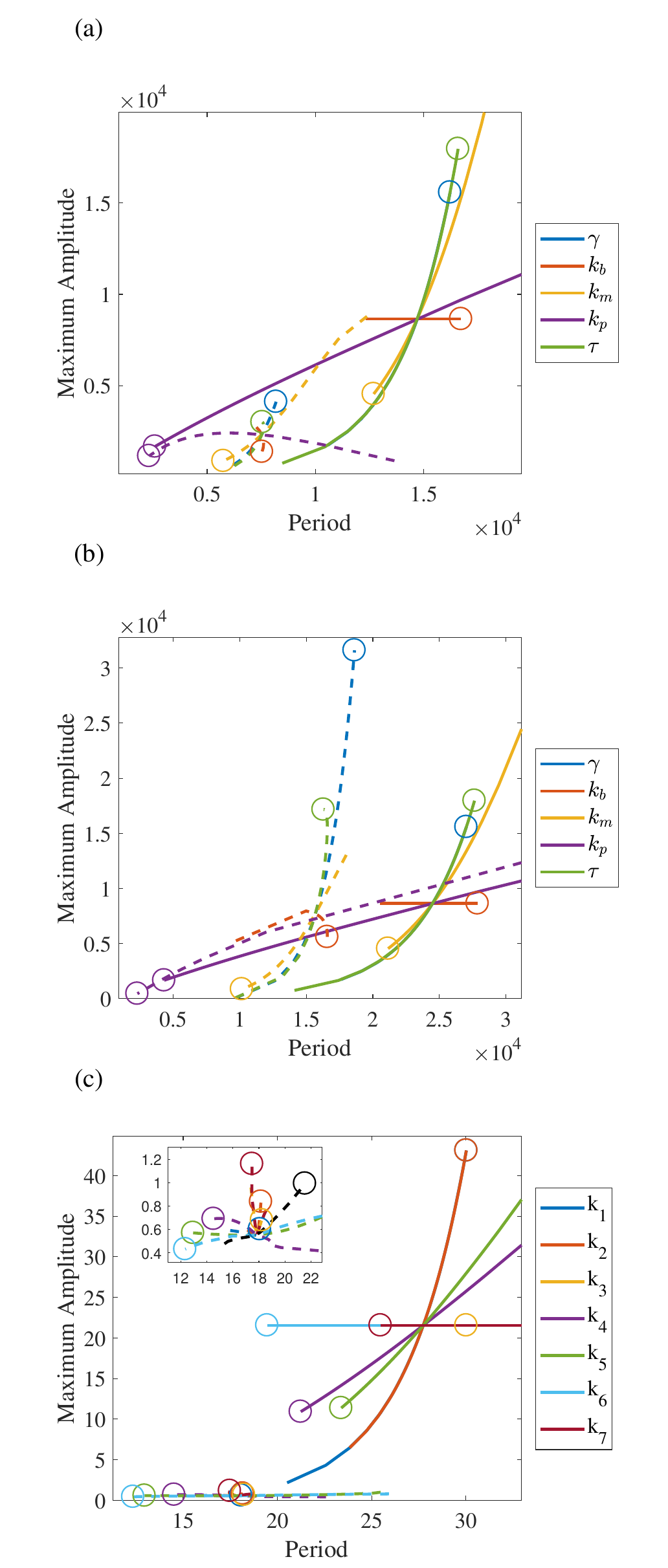}
	\caption{Analytical approximation of the co-variation between maximum amplitude and period. Dashed lines are for the numerical solutions of the ODE whereas  solid line curves for the analytical approximation for the period and maximum amplitude. Circle shaped markers represent the largest value of the corresponding parameter. a) Repressilator. b) Pentilator. c) Goodwin oscillator. Inset shows the zoomed version of the amplitude metric to period co-variation.} 
	\vspace{-.2in}
	\label{fig4}
\end{figure}

These approximate scaling laws are plotted in Fig. \ref{fig4}a  as parameters are 
varied individually. These show that the period and maximum amplitude increase or 
decrease simultaneously, except in cases where the amplitude is constant. Therefore, the main trend from these 
scaling laws is that as amplitude increases the period also increases.

We repeated the above exercise for the Pentilator and the Goodwin oscillator. Period ($T_{approx}$) and maximum amplitude ($A_{approx}$) approximations for Pentilator are given as,

\begin{equation}\label{penti_analytic}
\begin{aligned}
T_{approx}  & = 5\Big[2 \dfrac{\ln(2)}{k_m}+\dfrac{1}{k_p}\Big(\ln(X)+\ln\Big(\dfrac{X}{X-1}\Big)\Big)\Big], \\
A_{approx}  & = \dfrac{\tau \gamma}{k_p k_m},
\end{aligned}
\end{equation}  where $X = \dfrac{\tau \gamma \sqrt{k_b}}{k_p k_m}$.  

For Goodwin oscillator, the approximated period ($T_{approx}$) and amplitude ($A_{approx}$) are
\begin{equation}\label{gw_analytic}
\begin{aligned}
T_{approx}  & = 2\ln(2) [\dfrac{1}{k_4}+\dfrac{1}{k_5}]+\dfrac{1}{k_6}\Big[\ln(X)+\ln\Big(\dfrac{X}{X-1}\Big)\Big], \\
A_{approx}  & = \dfrac{k_1 k_2}{k_4 k_5},
\end{aligned}
\end{equation} 
where $X = \dfrac{k_1 k_2 k_3 }{k_4 k_5 k_6 k_7} $. 
\noindent Using these approximations, we obtained similar conclusions for the Pentilator (Fig. \ref{fig4}b) and the Goodwin oscillator (Fig. \ref{fig4}c).

\onecolumn

\begin{table}[H]
	\centering
		\caption{Co-variation of period and amplitude metrics} \label{tab1}
	\begin{subtable}[t]{0.48\textwidth}
		\renewcommand\thesubtable{1a}
		\caption{\footnotesize{For Repressilator, Pentilator, Goodwin oscillator, Van der Pol oscillator, Fitzhugh-Nagumo oscillator, Frzilator and Cyanobacterial circadian oscillator.}}
		\begin{tabular}{@{\extracolsep{\fill}}cccc}\toprule
			Oscillator/ & Maximum &	Period & Metric \\
			Parameters& 	 Amplitude & & $M_p$ \\	\midrule
			Repressilator &  &  & \\ \hline
			$\gamma$ & \cellcolor[gray]{0.7}Increases & \cellcolor[gray]{0.7}Increases & \cellcolor[gray]{0.7}Increases \\ \hline
			$k_b$ & Decreases & Complex$^a$ & Decreases \\ \hline
			$k_m$ & \cellcolor[gray]{0.7} Decreases & \cellcolor[gray]{0.7} Decreases  & \cellcolor[gray]{0.7}Decreases  \\ \hline
			$k_p$ & Complex$^a$ & Decreases  & Complex$^a$ \\ \hline
			$\tau$  & \cellcolor[gray]{0.7} Increases & \cellcolor[gray]{0.7} Increases & \cellcolor[gray]{0.7}Increases \\ \midrule
			Pentilator &  &  & \\ \hline
			$\gamma$ & \cellcolor[gray]{0.7} Increases & \cellcolor[gray]{0.7} Increases & \cellcolor[gray]{0.7} Increases \\ \hline
			$k_b$ & Complex$^a$ & Increases & Decreases \\ \hline
			$k_m$  & \cellcolor[gray]{0.7} Decreases  & \cellcolor[gray]{0.7} Decreases & \cellcolor[gray]{0.7} Decreases \\ \hline
			$k_p$ & \cellcolor[gray]{0.7} Decreases &  \cellcolor[gray]{0.7} Decreases& \cellcolor[gray]{0.7} Decreases  \\ \hline
			$\tau$ & \cellcolor[gray]{0.7}Increases & \cellcolor[gray]{0.7}Increases & \cellcolor[gray]{0.7} Increases \\ \midrule
			Goodwin oscillator &  &  & \\ \hline
			$k_1$ & Complex$^b$ & \cellcolor[gray]{0.7} Increases  & \cellcolor[gray]{0.7} Increases  \\ \hline
			$k_2$ & Increases & Constant   &  Increases  \\ \hline
			$k_3$  & Complex$^a$ &  Constant &  Increases \\ \hline
			$k_4$ & Increases & Decreases & Constant \\ \hline
			$k_5$ & Complex$^b$ & \cellcolor[gray]{0.7} Decreases & \cellcolor[gray]{0.7} Decreases  \\ \hline
			$k_6$  & \cellcolor[gray]{0.7} Decreases & \cellcolor[gray]{0.7} Decreases & \cellcolor[gray]{0.7} Decreases \\ \hline
			$k_7$ & Complex$^b$ & Constant & Increases \\ \hline
			$n$ & \cellcolor[gray]{0.7}Increases & \cellcolor[gray]{0.7}Increases & \cellcolor[gray]{0.7}Increases  \\ \midrule
			Van der Pol oscillator &  &  & \\ \hline
			$\mu$ & Constant & Increases & Constant  \\ \hline
			$\omega$ & Constant & Decreases  & Increases \\ \midrule
			\multicolumn{4}{l}{Fitzhugh-Nagumo oscillator} \\ \hline
			$\theta$ & Decreases  & Complex$^b$ & Decreases  \\ \hline
			$\gamma$ & Constant &  Complex$^b$  & Increases \\ \hline
			$\omega$ & Constant & Complex$^b$ & Increases \\ \hline
			$\phi$  & Constant & Decreases  & Constant \\ \midrule
			Frzilator &  &  & \\ \hline
			$\phi$  & Increases & Decreases & Increases \\ \hline
			$d_f$ & Constant & Constant & Decreases \\ \hline
			$k_c$ & \cellcolor[gray]{0.7}Decreases  & \cellcolor[gray]{0.7}Decreases  & \cellcolor[gray]{0.7}Decreases \\ \hline
			$d_c$ & \cellcolor[gray]{0.7} Increases & \cellcolor[gray]{0.7} Increases & Complex$^b$  \\ \hline
			$k_e$ & \cellcolor[gray]{0.7}Decreases & \cellcolor[gray]{0.7}Decreases & \cellcolor[gray]{0.7}Decreases \\ \hline
			$d_e$ & Constant & Constant &  Constant \\ \midrule
			\multicolumn{4}{l}{Cyanobacteria circadian oscillator}\\ \hline
			$k^0_{UT} $ & Constant  & Constant & Constant \\ \hline
			$k^0_{TD}$  & Constant & Constant & Constant \\ \hline
			$k^0_{SD}$  & Constant & Constant & Constant \\ \hline
			$k^0_{US}$ & Constant & Constant & Constant \\ \hline
			$k^0_{TU}$ & Increases & Constant  & Decreases \\ \hline
			$k^0_{DT}$ & Constant & Constant  & Constant \\ \hline
			$k^0_{DS}$ & \cellcolor[gray]{0.7} Decreases & \cellcolor[gray]{0.7} Decreases & \cellcolor[gray]{0.7} Decreases \\ \hline
			$k^0_{SU}$ & \cellcolor[gray]{0.7}Increases & \cellcolor[gray]{0.7} Increases & \cellcolor[gray]{0.7} Increases \\ \hline
			$k^A_{UT}$ & Increases & Decreases & Increases \\ \hline
			$k^A_{TD}$  & Increases &\cellcolor[gray]{0.7} Decreases & \cellcolor[gray]{0.7} Decreases \\ \hline
			$k^A_{SD}$ & \cellcolor[gray]{0.7}Increases & \cellcolor[gray]{0.7}Increases & \cellcolor[gray]{0.7}Increases  \\ \hline
			$k^A_{US}$ & \cellcolor[gray]{0.7}Decreases & \cellcolor[gray]{0.7}Decreases & \cellcolor[gray]{0.7}Decreases \\ \hline
			$k^A_{TU}$  & Decreases & Constant & Decreases \\ \hline			
			$k^A_{DT}$ & \cellcolor[gray]{0.7}Increases & \cellcolor[gray]{0.7}Increases & \cellcolor[gray]{0.7}Increases \\ \hline
			$k^A_{DS}$ & \cellcolor[gray]{0.7}Increases & \cellcolor[gray]{0.7}Increases & \cellcolor[gray]{0.7}Increases \\ \hline
		\end{tabular}{}	
	\end{subtable}%
\end{table}
\begin{table}[H]
	\centering
	\renewcommand\thetable{1}
	\begin{subtable}{0.48\textwidth}
	\renewcommand\thesubtable{1b}
		\caption{\footnotesize{For Cyanobacteria circadian oscillator, Metabolator, Mixed feedback oscillator, Meyer-Stryer model of calcium oscillations and Kim-Forger oscillator.}}
		\begin{tabular}[t]{@{\extracolsep{\fill}}cccc}\toprule
			Oscillator/ & Maximum &	Period & Metric \\
			Parameters& 	 Amplitude & & $M_p$ \\	\midrule	
			\multicolumn{4}{l}{Cyanobacteria circadian oscillator}\\ \hline
			$k^A_{SU}$ & \cellcolor[gray]{0.7}Decreases & \cellcolor[gray]{0.7}Decreases & \cellcolor[gray]{0.7}Decreases \\ \hline
			$K_{1/2}$ & \cellcolor[gray]{0.7}Decreases & \cellcolor[gray]{0.7}Decreases & \cellcolor[gray]{0.7}Decreases \\ \midrule
			Metabolator	 	&	  	&	  	&	 		\\	\hline
			$V_{gly}$ & Constant & Decreases & Increases \\ \hline
			$k_{TCA}$ & Decreases & Constant & Decreases \\ \hline
			$k_1$ & Complex$^a$ & Decreases & Increases \\ \hline
			$k_{m,1}$ & Constant & Constant & Constant \\ \hline
			$k_2$ & Complex$^a$ & Decreases & Complex$^a$ \\ \hline
			$k_{m,2}$ & Decreases & Constant & Constant \\ \hline
			$k_{Ack,f}$ & \cellcolor[gray]{0.7}Increases & \cellcolor[gray]{0.7}Increases & \cellcolor[gray]{0.7}Increases \\ \hline
			$k_{Ack,r}$ & Complex$^a$ & \cellcolor[gray]{0.7} Decreases & \cellcolor[gray]{0.7}Decreases \\ \hline
			$C$ & Constant & Constant & Constant \\ \hline
			$H^+$ & Constant & Constant & Constant \\ \hline
			$K_{eq}$ & Constant & Constant & Constant \\ \hline
			$k_3$ & Constant & Constant & Constant \\	\hline
			$HOAC_E$	 	&	 	Constant	 	&	 	Constant	 	&	 	Constant \\ \hline
			$K_{g,1}$	 	&	 	Complex$^a$	 	&	 	Decreases	 	&	 	Increases \\ \hline
			$n$	 	&	 	\cellcolor[gray]{0.7}Increases	 	&	 	\cellcolor[gray]{0.7}Increases	 	&	 	\cellcolor[gray]{0.7}Increases \\ \hline
			$K_{g,2}$	 	&	 	Complex$^a$	 	&	 	Increases	 	&	 	Constant \\ \hline
			$K_{g,3}$	 	&	 	Complex$^a$	 	&	 	Decreases	 	&	 	Increases \\ \hline
			$\alpha_0$	 	&	 	Complex$^a$	 	&	 	Constant	 	&	 	Constant \\ \hline
			$\alpha_1$	 	&	 	Complex$^a$	 	&	 	Increases	 	&	 	Decreases \\ \hline
			$\alpha_2$	 	&	 	Complex$^a$	 	&	 	Decreases	 	&	 	Constant \\ \hline
			$\alpha_3$	 	&	 	Complex$^a$	 	&	 	Decreases	 	&	 	Increases \\ \hline
			$k_d$	 	&	 	Complex$^a$	 	&	 	Decreases	 	&	 	Increases \\ \midrule
			Mixed feedback oscillator  &  &  & \\ \hline
			$\alpha$ & \cellcolor[gray]{0.7} Increases & \cellcolor[gray]{0.7}Increases  & \cellcolor[gray]{0.7} Increases  \\ \hline
			$\sigma$ & Complex$^a$ & Constant & Increases \\ \hline
			$\gamma_x$ & \cellcolor[gray]{0.7}Decreases & \cellcolor[gray]{0.7}Decreases & \cellcolor[gray]{0.7}Decreases \\ \hline
			$\gamma_y$  & Increases & Decreases  & Increases \\ \hline
			$\tau_y$ & \cellcolor[gray]{0.7}  Increases & \cellcolor[gray]{0.7}Increases & Constant \\ \midrule
			\multicolumn{4}{l}{Meyer-Stryer model of calcium oscillations} \\ \hline	
			$c_1$	 	&	 	Increases	 	&	 	\cellcolor[gray]{0.7}Decreases	 	&	 	\cellcolor[gray]{0.7}Decreases \\ \hline
			$c_2$	 	&	 	\cellcolor[gray]{0.7}Increases	 	&	 	\cellcolor[gray]{0.7}Increases	 	&	 	\cellcolor[gray]{0.7}Increases \\ \hline
			$c_3$	 	&	 	Constant	 	& \cellcolor[gray]{0.7}	 	Decreases	 	&	 	\cellcolor[gray]{0.7}Decreases \\ \hline
			$c_4$	 	&	 	Complex$^a$	 	&	 	Complex$^a$	 	&	 	Decreases \\ \hline
			$c_5$	 	&	 	\cellcolor[gray]{0.7}Increases	 	&	 	\cellcolor[gray]{0.7}Increases	 	&	 	\cellcolor[gray]{0.7}Increases \\ \hline
			$c_6$	 	&	 	\cellcolor[gray]{0.7}Decreases	 	&	 	\cellcolor[gray]{0.7}Decreases	 	&	 	\cellcolor[gray]{0.7}Decreases \\ \hline
			$c_7$	 	&	 	\cellcolor[gray]{0.7}Increases	 	&	 	\cellcolor[gray]{0.7}Increases	 	&	 	Decreases \\ \hline
			$K_1$	 	&	 	\cellcolor[gray]{0.7}Increases	 	&	 	\cellcolor[gray]{0.7}Increases	 	&	 	\cellcolor[gray]{0.7}Increases \\ \hline
			$K_2$	 	&	 	\cellcolor[gray]{0.7}Decreases	 	&	 	\cellcolor[gray]{0.7}Decreases	 	&	 	\cellcolor[gray]{0.7}Decreases \\ \hline
			$K_3$	 	&	 	\cellcolor[gray]{0.7}Increases	 	&	 	\cellcolor[gray]{0.7}Increases	 	&	 	\cellcolor[gray]{0.7}Increases \\ \hline
			$R$	 	&	 	Complex$^a$	 	&	 	Complex$^a$	 	&	 	Decreases \\ \midrule
			Kim-Forger model &  &  &  \\ \hline
			$k_1$ & \cellcolor[gray]{0.7}Increases & \cellcolor[gray]{0.7} Increases  & \cellcolor[gray]{0.7}Increases \\ \hline
			$k_2$ & \cellcolor[gray]{0.7}Increases & \cellcolor[gray]{0.7}Increases  &  \cellcolor[gray]{0.7}Increases \\ \hline
			$k_3$ & Decreases &  \cellcolor[gray]{0.7}Increases &  \cellcolor[gray]{0.7}Increases\\ \hline
			$k_4$ & Complex$^a$ & Decreases & Complex$^a$ \\ \hline
			$k_5$ & Complex$^a$ & \cellcolor[gray]{0.7}Decreases & \cellcolor[gray]{0.7}Decreases \\ \hline
			$k_6$ & Complex$^a$ & \cellcolor[gray]{0.7}Decreases & \cellcolor[gray]{0.7}Decreases \\ \hline
			$k_7$ & Increases & Decreases & Increases \\ \hline
			$k_8$ & Constant & Constant & Constant \\ \hline
			\multicolumn{4}{l}{$^a$ First increases then decreases.}\\
			\multicolumn{4}{l}{$^b$ First decreases then increases.} \\
			\multicolumn{4}{l}{Constant: variations within $\pm 5\%$ from nominal values.}\\   
		\end{tabular}{}
	\end{subtable}%
\end{table}
\clearpage

\begin{table}[H]
	\renewcommand\thetable{2}
	\centering
	\caption{Categorization of Co-variation in period and amplitude} \label{tab2}
	\begin{subtable}[t]{0.45\textwidth}
		\renewcommand\thesubtable{2a}
		\caption{\footnotesize{For Repressilator, Pentilator, Goodwin oscillator, Van der Pol oscillator, Fitzhugh-Nagumo oscillator, Frzilator and Cyanobacterial circadian oscillator.}}
		\begin{tabular}[t]{@{\extracolsep{\fill}}lcc}\toprule
			\multicolumn{1}{c}{Oscillator/}  & Maximum&Metric \\ 
			\multicolumn{1}{c}{Parameters}&  amplitude &  $M_p$\\\midrule
			{Repressilator} & &\\ \hline
			$\gamma$, production rate constant & Type 1 & Type 1 \\ \hline
			$k_b$, dissociation constant  & Type 5 & Type 5\\ \hline
			$k_m$, degradation rate constant & Type 1 & Type 1 \\ \hline
			$k_p$, degradation rate constant & Type 5 & Type 5 \\ \hline
			$\tau$, production rate constant & Type 1 & Type 1 \\  \midrule
			{Pentilator} &  &\\ \hline
			$\gamma$, production rate constant & Type 1 & Type 1 \\ \hline
			$k_b$, dissociation constant  & Type 5 & Type 2 \\ \hline
			$k_m$, degradation rate constant  & Type 1 & Type 1 \\ \hline
			$k_p$, degradation rate constant  & Type 1 & Type 1 \\ \hline
			$\tau$, production rate constant  & Type 1 & Type 1 \\ \midrule
			{Goodwin oscillator} &  & \\ \hline
			$k_1$, production rate constant & Type 5 & Type 1 \\ \hline
			$k_2$, production rate constant & Type 4 & Type 4 \\ \hline
			$k_3$, production rate constant  & Type 5 & Type 4 \\ \hline
			$k_4$, degradation rate constant  & Type 2 & Type 3 \\ \hline
			$k_5$, degradation rate constant & Type 5 & Type 1 \\ \hline
			$k_6$, degradation rate constant & Type 1 & Type 1 \\ \hline
			$k_7$, dissociation constant  & Type 5 & Type 2 \\ \hline
			$n$, Hill coefficient & Type 1 & Type 1 \\ \midrule
			{Van der Pol oscillator} & &\\ \hline
			$\mu$ & Type 3 & Type 3 \\ \hline
			$\omega$ & Type 3 & Type 2 \\ \midrule
			\multicolumn{3}{l}{Fitzhugh-Nagumo oscillator} \\ \hline
			$\theta$ & Type 5 & Type 5 \\ \hline
			$\gamma$ & Type 5 & Type 5 \\ \hline
			$\omega$ & Type 5 & Type 5 \\ \hline
			$\phi$   & Type 3 & Type 3 \\ \midrule
			{Frzilator} & & \\ \hline
			$\phi$, production rate constant & Type 2 & Type 2\\ \hline
			$d_f$, degradation rate constant & Type 0 & Type 4\\ \hline
			$k_c$, production rate constant  & Type 1 & Type 1\\ \hline
			$d_c$, degradation rate constant & Type 1 & Type 5 \\ \hline
			$k_e$, production rate constant  & Type 1 & Type 1\\ \hline
			$d_e$, degradation rate constant & Type 0 & Type 0\\ \midrule
			\multicolumn{3}{l}{Cyanobacteria circadian oscillator} \\ \hline
			$k^0_{UT}$, inter conversion rate & Type 0 & Type 0\\ \hline
			$k^0_{TD}$, inter conversion rate  & Type 0 & Type 0\\ \hline
			$k^0_{SD}$, inter conversion rate  & Type 0 & Type 0\\ \hline
			$k^0_{US}$, inter conversion rate & Type 0 & Type 0\\ \hline
			$k^0_{TU}$, inter conversion rate  & Type 4 & Type 4\\ \hline
			$k^0_{DT}$, inter conversion rate & Type 0 & Type 0\\ \hline
			$k^0_{DS}$, inter conversion rate & Type 1 & Type 1\\ \hline
			$k^0_{SU}$, inter conversion rate & Type 1 & Type 1\\ \hline
			$k^A_{UT}$, inter conversion rate & Type 2 & Type 2\\ \hline
			$k^A_{TD}$, inter conversion rate & Type 2 & Type 1\\ \hline
			$k^A_{SD}$, inter conversion rate & Type 1 & Type 1\\ \hline
			$k^A_{US}$, inter conversion rate & Type 1 & Type 1\\ \hline
			$k^A_{TU}$, inter conversion rate & Type 4 & Type 4\\ \hline
			$k^A_{DT}$, inter conversion rate & Type 1 & Type 1\\ \hline
		\end{tabular}{}	
	\end{subtable}%
\end{table}
\begin{table}[H]
	\centering
	\renewcommand\thetable{2}
	\begin{subtable}[t]{0.45\textwidth}
		\flushright
		\renewcommand\thesubtable{2b}
		\caption{\footnotesize{For Cyanobacteria circadian oscillator, Metabolator, Mixed feedback oscillator, Meyer-Stryer model of calcium oscillations and Kim-Forger oscillator.}}
		\begin{tabular}[t]{@{\extracolsep{\fill}}lcc}\toprule
			\multicolumn{1}{c}{Oscillator/}  & Maximum&Metric \\ 
			\multicolumn{1}{c}{Parameters}&  amplitude &  $M_p$\\\midrule
			\multicolumn{3}{l}{Cyanobacteria circadian oscillator} \\ \hline	
			$k^A_{DS}$, inter conversion rate & Type 1 & Type 1\\ \hline
			$k^A_{SU}$, inter conversion rate & Type 1 & Type 1\\ \hline
			$K_{1/2}$, dissociation constant & Type 1 & Type 1\\ \midrule
			{Metabolator} & & \\ \hline
			$V_{gly}$, production rate & Type 3 & Type 2\\ \hline
			$k_{TCA}$, degradation rate constant & Type 4 & Type 4\\ \hline
			$k_1$ & Type 5 & Type 2\\ \hline
			$k_{m,1}$ & Type 0 & Type 0\\ \hline
			$k_2$ & Type 5 & Type 5\\ \hline
			$k_{m,2}$ & Type 4 & Type 0\\ \hline
			$k_{Ack,f}$ & Type 1 & Type 1\\ \hline
			$k_{Ack,r}$ & Type 5 & Type 1\\ \hline
			$C$ & Type 0 & Type 0\\ \hline
			$H^+$ & Type 0 & Type 0\\ \hline
			$K_{eq}$ & Type 0 & Type 0\\ \hline
			$k_3$, degradation rate constant & Type 0 & Type 0 \\ \hline
			$HOAC_E$ & Type 0 & Type 0\\ \hline
			$K_{g,1}$, dissociation constant& Type 5 & Type 2\\ \hline
			$n$, Hill coefficient & Type 1 & Type 1\\ \hline
			$K_{g,2}$, dissociation constant & Type 5 & Type 3\\ \hline
			$K_{g,3}$, dissociation constant & Type 5 & Type 2\\ \hline
			$\alpha_0$, leaky coefficient & Type 0 & Type 0\\ \hline
			$\alpha_1$, production rate constant & Type 5 & Type 2\\ \hline
			$\alpha_2$, production rate constant & Type 5 & Type 4\\ \hline
			$\alpha_3$, production rate constant & Type 5 & Type 2\\ \hline
			$k_d$, degradation rate constant & Type 5 & Type 2\\ \midrule
			{Mixed feedback oscillator}  &  & \\ \hline
			$\alpha$ & Type 1 & Type 1 \\ \hline
			$\sigma$ & Type 4 & Type 4 \\ \hline
			$\gamma_x$, degradation rate constant & Type 1 & Type 1\\ \hline
			$\gamma_y$, degradation rate constant & Type 2 & Type 2\\ \hline
			$\tau_y$ & Type 1 & Type 3\\ \midrule
			\multicolumn{3}{l}{Meyer-Stryer model of calcium oscillations} \\ \hline
			$c_1$ & Type 2 & Type 1\\ \hline
			$c_2$ & Type 1 & Type 1\\ \hline
			$c_3$ & Type 3 & Type 1\\ \hline
			$c_4$ & Type 5 & Type 5\\ \hline
			$c_5$, degradation rate constant & Type 1 & Type 1\\ \hline
			$c_6$ & Type 1 & Type 1\\ \hline
			$c_7$ & Type 1 & Type 2\\ \hline
			$K_1$, dissociation constant & Type 1 & Type 1\\ \hline
			$K_2$, dissociation constant & Type 1 & Type 1\\ \hline
			$K_3$, dissociation constant & Type 1 & Type 1\\ \hline
			$R$, production rate constant & Type 5 & Type 5\\ \midrule
			{Kim-Forger model} & & \\ \hline
			$k_1$, production rate constant & Type 1 & Type 1\\ \hline
			$k_2$, production rate constant & Type 1 & Type 1\\ \hline
			$k_3$, production rate constant & Type 2 & Type 1\\ \hline
			$k_4$, degradation rate constant & Type 5 & Type 5\\ \hline
			$k_5$, degradation rate constant & Type 5 & Type 1\\ \hline
			$k_6$, degradation rate constant & Type 5 & Type 1\\ \hline
			$k_7$ & Type 2 & Type 2\\ \hline
			$k_8$, dissociation constant & Type 0 & Type 0\\ 
		\end{tabular}{}
	\end{subtable}
\end{table}

\section{Discussion}\label{sec3}

Understanding design principles underlying operation of biomolecular oscillators is an important challenge. Using computational models of benchmark oscillators --- Repressilator, Pentilator, Goodwin oscillator, Van der Pol oscillator, Fitzhugh-Nagumo oscillator, Frzilator, Cyanobacterial circadian oscillator, Metabolator, a mixed feedback oscillator and Meyer-Stryer model of calcium oscillations and Kim-Forger model --- we present an investigation of the co-variation of period and amplitude. First, we plot the maximum amplitude versus the period as each parameter is individually varied in a range around the nominal value, finding a range of curves showing both mutually increasing as well as other trends. Second, noting that the maximum amplitude metric considers a single state, we adapted a metric based on the power of a multivariable output signal and plotted these against the period, finding largely similar co-variation trends. Third, we analytically approximate three oscillator models to obtain "scaling laws" for the period-amplitude co-variation, finding that as approximate period increases the upper bound of maximum amplitude always increases (or remains constant). These results provide an insight into the functioning of biomolecular oscillators as well as a  reference to choose specified amplitude-period values for design applications.

We note how the scaling laws obtained show that, throughout the parameter space where oscillations persist, as the approximation to the period increases, the upper bound on the maximum amplitude either increases or, due to saturation effects, reaches a constant.
This is consistent with a Type 1, or Type 3 behaviour.
This analysis also underlines that the region in period-amplitude space where the period is low and amplitude is high is hard to access, denoted `Forbidden Region' in Fig. \ref{fig1}.
The actual co-variation of the period and amplitude as a function of the variation of the parameters may have more complex patterns away from the Forbidden Region. Based on the computations presented above, we discuss these patterns (Table \ref{tab2}) below.

\textit{Production Rate:} The production rate controls the extent of the production of the protein.
As this increases, we expect the maximum amplitude to increase.
Further, other parameters being equal, we expect an increase in the period on account of the extra time that may be required for the biomolecules to reach the higher amplitude levels.
The simultaneous increase in period and amplitude should classify this as Type 1 behaviour, or a Type 3 behaviour in case saturation effects dominate. 
For the Repressilator, we found that the production rate constants $\gamma$ and $\tau$ exhibited a Type 1 co-variation between the period and the maximum amplitude. 
Similar trends were observed for the Pentilator.
However, for other cases, such as the production rate constants $k_1$ and $k_3$ in the Goodwin oscillator, other types of co-variations are observed.
This could be because of dual dynamical effects of the production rate.
For example, an increase in production rate may trigger a downstream repressor that may reduce the maximum amplitude.
Out of the 18 instances when the parameter can be classified as a production rate constant, there were 9 instances of the production rate constant exhibiting a Type 1 or Type 3 behaviour in the co-variation of the period and the maximum amplitude.

\textit{Degradation Rate:}
The degradation rate constant is a parameter that controls the rate at which a biomolecule degrades.
We expect that as this increases, the maximum amplitude possible should decrease.
Further, we expect that the period should also decrease, both from the point of view of the lower amplitudes that needs to be reached as well as from the point of view the effect on the time constant of the system.
The time constant is inversely proportional to the degradation rate constant, and as this rate increases, the time constant reduces, or response speeds up.
Again, this is an example of a Type 1 behaviour as both the maximum amplitude and period simultaneously decrease.
For the Repressilator, we found that the degradation rate constant $k_m$ exhibited a Type 1 co-variation between the period and the maximum amplitude.
 However, for the degradation constant $k_p$, a complex co-variation pattern (Type 5) was observed, possibly due to a dual effect of the parameter. For Pentilator, on the other hand, both the degradation rate constants exhibited Type 1 behaviour.
For other oscillators, among 15 instances of degradation rate constant, there were 5 instances of Type 1 co-variation between the period and the maximum amplitude.

\textit{Dissociation Constant:}
The dissociation constant reflects the rate at which the activation/ repression of a biomolecule is at half its maximum rate.bAs this increases, we expect that the amplitude also increases, owing to the need for biomolecules to reach higher levels to trigger activation/ repression.
We expect that reaching higher amplitude may take more time and so the period should increase. Therefore, these expectations are of a Type 1 behaviour.
 
However, we found that the dissociation constant ($1/k_b$) in the Repressilator and the Pentilator as well as the dissociation constant $k_7$ in the Goodwin oscillator exhibited complex co-variation (Type 5) behaviour.Perhaps in some parameter regime, there is a dual effect of this parameter.

Finally, we note how different mechanisms underlying circadian oscillations can have a similar period-amplitude co-variation. In the context of circadian clocks, the negative feedback may be implemented by a protein sequestration-based mechanism as well as a Hill-type repression mechanism, as in the Kim-Forger model and the Goodwin model, respectively~\cite{kimforger, goodwin}.
The key difference between the two models is the production term for the mRNA, which is either a Hill-type repression or a protein sequestration-based mechanism.
When we compare period-amplitude co-variations of these two models, for the nominal parameter sets, we find certain similarities as well as differences.
On the one hand, the quantitative nature of the co-variations are different, with the same parameter giving different trends.
For example, as the degradation rate constant $k_5$ is increased, the period decreases in both models.
However, for the Goodwin oscillator, the maximum amplitude first decreases, then increases, whereas for the Kim-Forger oscillator the opposite happens (Please see Supplementary material: Figs. 3B, 11B).
On the other hand, on a closer look, we find that both models can have the property that the amplitude can be tuned while the period does not change significantly using same parameters ($k_2$, $k_3$, $k_7$).
Similarly, both models can have the property that the period, and to some extent the amplitude, can be tuned using same parameters ($k_4$, $k_5$, $k_6$), all of which are related to the degradation reaction.
Further, the scaling laws for the Kim-Forger oscillator are similar to that of the Goodwin oscillator if the upper bound for the amplitude is assumed conservatively.
This is consistent with the presence of a `forbidden region' indicating that achieving large amplitude small period oscillations may be difficult.
These similarities could be because of the fact that, despite different biomolecular mechanisms, they have similar mathematical representations at the systems-level. It also points to similar constraints that may have shaped their evolution.

\section*{Acknowledgments}\label{sec11}

We thank the anonymous reviewers for their helpful comments. We thank Prof. I. N. Kar, Prof. S. Janaradhanan and Mr. Abhilash Patel for their valuable inputs. This research is supported
partially by Science and Engineering Research Board grant SB/FTP/ETA-0152/2013.

\bibliographystyle{iet}
\bibliography{ietreflist}
\end{document}